# Collision and Preimage Resistance of the Centera Content Address

Robert Primmer †,  Carl D'Halluin ‡

**Abstract**

*Centera uses cryptographic hash functions as a means of addressing stored objects, thus creating a new class of data storage referred to as CAS (content addressed storage). Such hashing serves the useful function of providing a means of uniquely identifying data and providing a global handle to that data, referred to as the Content Address or CA. However, such a model begs the question: how certain can one be that a given CA is indeed unique?*

*In this paper we describe fundamental concepts of cryptographic hash functions, such as collision resistance, preimage resistance, and second-preimage resistance. We then map these properties to the MD5 and SHA-256 hash algorithms, which are used to generate the Centera content address. Finally, we present a proof of the collision resistance of the Centera Content Address.*

## 1   Introduction

EMC introduced Centera in April 2002, ushering in a new model of data storage referred to as Content Addressed Storage, or CAS [8]. The two principal ideas Centera introduced were a node-based structure referred to as RAIN, for redundant array of independent nodes, and an object store.

Centera presents a flat object store to the application. Applications submit data and are returned a global handle that uniquely refers to the data stored. The handle returned is a function of the data submitted. Upon submission, a 256-bit handle — called the Content Address, or CA — is calculated and returned to the application [20]. Nodes are divided into two groups: *Access Nodes*, which act principally as gateways between the cluster and the client application server; and *Storage Nodes*, which act principally as semi-autonomous storage containers. Data objects are referred to as *blobs*, and the associated metadata containers are referred to as *clips*.

The CA serves the twin role of unique data handle and MDC (Manipulation Detection Code). It is the hashing function that allows for the latter role. Since the CA is calculated based upon a bit representation of the content itself, any modification of the content results in a different CA. Centera takes advantage of this property for achieving cache coherence and detecting when a file on disk has been damaged or in some way altered. Detail of this mechanism and the RAIN structure can be found in the Centera Product Description Guide [9].

In this paper we provide a general description of hash functions and their properties. We then focus on the MDx and SHA-x family of hash functions as the MD5 and SHA-256 schemes are currently employed by Centera for a portion of the CA calculation. We then describe the most common forms of attack on hash functions and relate those specifically to the Centera addressing schemes. Finally we present a proof of the collision and preimage resistance of the full CA used by Centera as the handle to stored clips and blobs.

## 2   Hash Functions

A hash function maps bit strings of an arbitrary finite length to fixed-length hash values. On the one hand a hash function must be fast to compute, but on the other it must also be collision-resistant, i.e. it must be computationally infeasible to find a collision (that is a pair of messages with the same hash value) [11]. Additionally, hash functions are intended to resist inversion (i.e. finding a message from a given hash value).

The basic idea of cryptographic hash functions is that a hash value serves as a compact representative image (sometimes called a *message digest*) of an input string, and can be used to uniquely identify that string. There are several types of hash functions [6][25]. The best known are:

- Hash functions based on block ciphers;
- Hash functions based on modular arithmetic;
- Hash functions based on cellular automata;
- Hash functions based on knapsack problem;
- Hash functions based on algebraic matrices; and
- Dedicated hash functions.

In this paper we focus on dedicated hash functions. Readers interested in a non-mathematical treatment of hash functions and cryptography are referred to [28] and [27] respectively.

### 2.1   Definitions

Perhaps the most widely known definitions for collision resistance, preimage-resistance, and second-preimage resistance come from the Handbook of Applied Cryptography (HAC) [2] which is based in part on the work of Preneel [5] which is the seminal work in this area, providing extensive treatment of cryptographic hash functions according to the information theoretic, complexity theoretic, and system based approaches. For our purposes we will rely primarily upon his descriptions.

† EMC² Corporation, Hopkinton, MA USA

‡ EMC² Corporation, Mechelen, Belgium

The following terms are basic to cryptography in general and hash functions in particular.

Definition 1: A **one-way hash function** (OWHF) is a function *h* satisfying the following conditions:

1. *The argument X can be of arbitrary length and the result h(X) has a fixed length of n bits.*

2. *Given h and X, the computation of h(X) must be "easy".*

3. *The hash function must be one-way in the sense that given a Y in the image of h, it is "hard" to find a message X such that h(X) = Y (preimage resistant) and given X and h(X) it is "hard" to find a message $X' \neq X$ such that $h(X') = h(X)$ (second-preimage resistant).*

Definition 2: A **collision resistant hash function** (CRHF) is a one-way hash function with the additional property that the hash function must be collision resistant: this means that it is "hard" to find two distinct messages that hash to the same result. That is, given any two distinct inputs $X, X'$ it is computationally infeasible to hash to the same output such that $h(X) = h(X')$.

Note that "preimage resistant" is alternatively referred to as *one-way*; "(second) preimage resistant" is referred to as *weak collision resistant*; and "collision resistant" is referred to as *strong collision resistant*. Further mathematical rigor to the definition of these terms can be found in [2] and [17].

Definition 3: An *n*-bit hash function has **ideal security** if it satisfies the following conditions:

1. *Given a hash output, producing a preimage or second preimage requires ~$2^n$ operations.*

2. *Producing a collision requires ~$2^{n/2}$ operations.*

Definition 4: **Manipulation detection code** (MDC) is the umbrella term used to describe the class of hash functions that are unkeyed.

The tradeoff that must be considered when choosing a specific algorithm in this class is the tension between efficiency of computation versus degree of security. The two are often inversely related. Because the speed of a hash function is typically of critical importance, many block ciphers – while providing excellent security - are too slow to be practical. This gave rise to specialized hash functions, which are designed from scratch with speed as a principal consideration.

Definition 5: **Message authentication code** (MAC) is the umbrella term used to describe the class of hash functions that are keyed. While the introduction of a key adds state, it has the positive effect of allowing for expanded use of the hash function, with applications in areas such as digital signature authentication. For example, Centera uses a keyed MAC to validate operations between management station(s) and the Centera server.

A MAC is a function which takes the secret key *k* (shared between Alice and Bob) and the message *m* to return a tag $MAC_k(m)$. Here, Eve represents the adversary (eavesdropper). Eve sees a sequence $(m_1,a_1)$, $(m_2,a_2)$,…,$(m_q,a_q)$ of pairs of messages and their corresponding tags (i.e., $a_i = MAC_k(m_i)$) transmitted between the parties. Eve breaks the MAC if she can find a message *m*, not included among $m_1,m_2,…,m_q$, together with its corresponding valid authentication tag $a = MAC_k(m)$. Eve's success probability is the probability that she breaks the MAC.

Hash functions are generally not keyed. However, there is interest in using cryptographically strong hash functions as the basis for MAC schemes. Bellare et al. [16] describe a means of using an iterative hash function as the basis for a MAC by keying the hash function via the initial vector, instead of the more common mechanism of inputting the key as part of the data hashed by the function.

## 2.2 Iterated Hash Functions

Most known hash functions (including MD5 and SHA) are based on a compression function with fixed size input; they process every message block in a similar way. The information is divided into *t* blocks $X_1$ through $X_t$. If the total number of bits is not a multiple of the block length, the information is padded to the required length (using a so-called *padding rule*). The hash function can then be described as follows:

$$H_0 = IV$$
$$H_i = f(X_i, H_{i-1}) \quad i = 1, 2, ..., t \quad (2.1)$$
$$h(X) = g(H_t)$$

Where *IV* is the abbreviation for Initial Value (or Initial Vector); $H_{i-1}$ serves as the *n*-bit chaining variable between stage *i* – 1 and stage *i*; and the result of the hash function is denoted with *h(X)*. The function *f* is called the *round* or *compression* function, and the function *g* is called the *output transformation*. It is often omitted (that is, *g* is often the identity function); in the case of Centera, *g* maps bit strings to a base32-notation (digits 0-9 and A-V), which has no impact on the security of the hash. Two elements in this definition have an important influence on the security of a hash function: the choice of the padding rule and the choice of the *IV*. Figure 1 illustrates the mechanics of a typical iterated hash function [2].

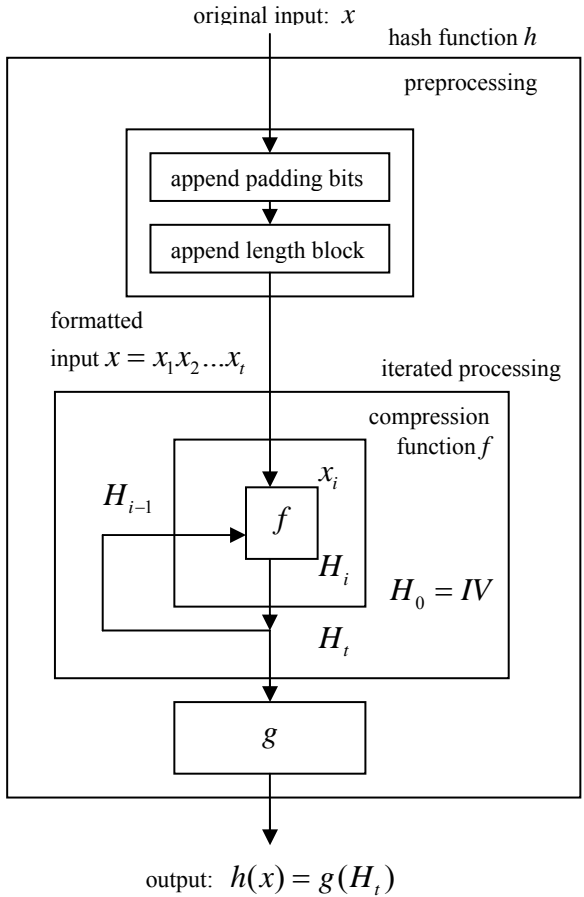

output: $h(x) = g(H_t)$

**Figure 1. Iterated Hash Functions**

It is recommended that the padding rule is unambiguous (i.e., there do not exist two messages that can be padded to the same padded message); at the end one should append the length of the message; and the *IV* should be defined as part of the description of the hash functions (this is called MD-strengthening, after Merkle and Damgård). In some cases one can deviate from this rule, but this will make the hash function less secure and may lead to trivial collisions or second preimages. Note that the function *f*, the *IV*, and the padding scheme are publicly known.

Research on hash functions has been focused on the question: which properties should be imposed on *f* to guarantee that *h* satisfies certain properties? Two partial answers have been found to this question. We state them here for the case that *g* is the identity function. The first result is by Lai and Massey [29] and gives necessary and sufficient conditions for *f* in order to obtain an "ideally secure" hash function *h*.

**Theorem 1 (Lai-Massey)** *Assume that the padding contains the length of the input string, and that the message X (without padding) contains at least 2 blocks. Then finding a second-preimage for h with a fixed IV requires $2^n$ operations if and only if finding a second-preimage for f with arbitrarily chosen $H_{i-1}$ requires $2^n$ operations.*

The fact that the condition is necessary is based on the following argument: if one can find a second-preimage for *f* in $2^s$ operations (with $s < n$), one can find a second-preimage for *h* in $2^{1+(n+s)/2}$ operations with a meet-in-the-middle attack [5].

A second result by Damgård [15] and independently by Merkle [24] states that for *h* to be a CRHF it is sufficient that *f* is a collision resistant function.

**Theorem 2 (Damgård-Merkle)** *Let f be a collision resistant function mapping l to n bits (with l – n > 1). If an unambiguous padding rule is used, the following construction yields a CRHF:*

$$H_1 = f(0^{n+1} \| X_1)$$
$$H_i = f(H_{i-1} \| 1 \| X_i) \quad i = 2,3,...,t \quad (2.2)$$
$$h(X) = H_{t+1}$$

Where $\|$ denotes concatenation of bit strings and $0^{n+1}$ indicates a sequence of $n+1$ bits, all being zero.

The proof that the resulting function *h* is collision resistant follows by a simple argument that a collision for *h* would imply a collision for *f* for some stage *i*. The inclusion of the length-block, which effectively encodes all messages such that no encoded input is the tail end of any other encoded input, is necessary for this reasoning [2].

Recent findings [11] show that for a typical *n*-bit iterated hash function, one can find a second preimage for a $2^k$-block message in about $k \cdot 2^{n/2+1} + 2^{n-k+1}$ operations for long messages, assuming unlimited memory.

## 3  MDx and SHA Functions

The MDx functions are a form of "Dedicated hash functions" defined earlier in §2. The MD4 [21] hash function was introduced by Ron Rivest in 1990. While weaknesses were discovered with MD4 [15][30][32], MD4's significance is that it served as the starting point for the development of a series of similar iterated hash functions. MD4-derived hash functions include the 256-bit extension of MD4 [23] and the Secure Hash Algorithm (SHA) designed by NIST/NSA, later substituted by the slightly revised SHA-1 [10]. Additionally MD4 is optimized to be highly efficient on 32-bit word computers and is thus able to produce 128-bit hash results far quicker than alternative methods, e.g. via modified block ciphers.

To counter the weaknesses discovered with MD4 Rivest introduced MD5 [22] in 1991 to strengthen the hash function accordingly. MD5 differs from MD4 on the following points [3].

- A fourth round was added.

- The second round function has been changed from the majority function $(X \wedge Y) \vee (X \wedge Z) \vee (Y \wedge Z)$ to the multiplexer function $(X \wedge Z) \vee (Y \wedge \neg Z)$.
- The order in which input words are accessed in rounds 2 and 3 was changed.
- The shift amounts in each round have been changed. None are the same now.
- Each step now has a unique additive constant.
- Each step now adds in the result of the previous step.

Where $\wedge$ denotes bit-wise AND, $\vee$ denotes bit-wise OR, $\oplus$ denotes bit-wise XOR, and $\neg Z$ denotes the bit-wise complement of $Z$.

## 3.1  MD5 Detail

MD5 is perhaps the most extensively covered dedicated hash function in the literature, likely due to its popularity. Rivest details the algorithm and provides sample C source in [22]. Schneier provides a survey of several of the more frequently used hash functions in [7]. For our purposes a subset of that information is sufficient.

The MD5 algorithm follows the iterative structure described in §2.2, where the hash is computed by repeated application of a compression (or *round*) function to successive blocks of the message. The message to be hashed is first padded to a multiple of 512 bits and then divided into a sequence of 512-bit message blocks. The compression function takes two inputs, a 128-bit chaining value and a 512-bit message block, and produces as output a new 128-bit chaining variable, which is input to the next iteration of the compression function [4]. In every step one of the chaining variables is updated. A typical step operation of the round function is

$$A = (B + ((A + \Phi(B,C,D) + X + K) << s),$$

where $X$ is the 512-bit input block (split into $i$ 32-bit words), + denotes addition modulo $2^{32}$, $\Phi$ is the round-dependent Boolean function (e.g. XOR), $K$ is the step-dependent constant, and $<< s$ denotes a left circular shift by $s$ positions [12].

Rivest defined four 32-bit variables (referred to as chaining variables) that are initialized to the following values:

$$\begin{align}
\texttt{A} &= \texttt{0x01234567} \\
\texttt{B} &= \texttt{0x89abcdef} \\
\texttt{C} &= \texttt{0xfedcba98} \\
\texttt{D} &= \texttt{0x76543210}
\end{align} \quad (3.1)$$

There are four nonlinear functions, one used in each operation (a different one for each round).

$$\begin{align}
\Phi_1(X,Y,Z) &= (X \wedge Y) \vee (\neg X \wedge Z) \\
\Phi_2(X,Y,Z) &= (X \wedge Z) \vee (Y \wedge \neg Z) \\
\Phi_3(X,Y,Z) &= X \oplus Y \oplus Z \\
\Phi_4(X,Y,Z) &= Y \oplus (X \vee \neg Z)
\end{align} \quad (3.2)$$

These functions are designed so that if the corresponding bits of $X$, $Y$, and $Z$ are independent and unbiased, then each bit of the result will also be independent and unbiased. The function $\Phi_1$ is the bit-wise conditional: If $X$ then $Y$ else $Z$. The function $\Phi_2$ is the bit-wise conditional: If $Z$ then $X$ else $Y$. The function $\Phi_3$ is the bit-wise parity operator [7]. After the last message block has been processed, the final chaining value is output as the hash of the message.

## 3.2  SHA Functions

The Secure Hash Algorithm (SHA) [10] is a digest algorithm proposed by the US NIST (National Institute of Standards and Technology) agency as a standard digest algorithm in August 2002. NIST released a first version of this algorithm — referred to as "SHA" back in 1992. It subsequently discovered a weakness with SHA and released a second version which it referred to as "SHA-1".

Where SHA-1 has a 160 bit hash value, NIST also defined hash functions with larger hash output such as SHA-224 (224 bits), SHA-256 (256 bits), SHA-384 (384 bits) and SHA-512 (512 bits). The cryptographic community generally considers SHA-256, SHA-384 and SHA-512 to be stronger than MD5.

Detail on the internals of the SHA family can be found in [10].

## 4  Hash Collisions

MD5 satisfies the necessary conditions to be classified as a OWHF. As such it is widely used in security protocols such as TLS and IPsec. Perhaps due to its speed of computation, MD5 is among the most frequently used hash functions in use today. However, researchers have forged collisions in a subset of the round functions in MD5 [3][12]. Moreover, at *Crypto 2004* a paper was presented [30] that demonstrated artificially constructed MD5 hash collisions between two 1024-bit streams having a 6 bit difference. More details on the methodology are available in [31].

Researchers have also found a methodology [33] to induce collisions in a reduced-round version of SHA-1. However, since it still requires a supercomputer to calculate a colli-sion for SHA-1, no practical collisions are known presently.

SHA-256, SHA-384 or SHA-512 are considered very secure and no attacks are known presently.

## 4.1 Practical Effects of Induced Collisions

It is important to contrast the results of theoretical work, where collisions can be constructed based on specially crafted inputs that differ in a small number of bits, versus the use of a hash function in practice on real data.

For practical use in a content addressed system such as Centera, the more important concern in the ordinary course of usage is how likely it will be that two distinct messages (files in this case), will actually resolve to the same hash value. This is different from the results for specially crafted inputs that are intended to have the same hash value. As the next section describes, the probability of naturally occurring collisions is exceedingly small.

## 4.2 Non-induced Collisions

As stated in §2.1 Definition 2, for a hash function to be strong collision resistant it must be one-way (i.e. both preimage and second-preimage resistant) and it must be hard to find two distinct messages that hash to the same result. As used in Centera, in the worst case a preimage or second-preimage for a given file would not be found on average until $\approx 2^{127}$ (170 billion billion billion billion) unique files were stored on a single logical cluster (refer to Theorem 1). Producing a hash collision between any two distinct files using just MD5 would not be found on average until $\approx 2^{64}$ (18 billion billion) unique files were stored on a single logical cluster (refer to formula (5.2)).

The next sections apply the properties of hash collisions to the three naming schemes used by Centera.

## 5 Collision Probabilities of Centera CA

Centera has three naming schemes, referred to as **M**, **GM** and **M++**. The **M** naming scheme returns a 128-bit CA that is simply the MD5 hash of the associated content. The **GM** naming scheme returns a 256-bit CA that is comprised of an MD5 hash of the associated content, a random bit-string, timestamp, counter and header bits. The **M++** naming scheme returns a 256-bit CA that is comprised of the MD5 and a truncated SHA-256 hash of the associated content. Detail for **M** is provided in §5.2; **GM** detail is provided in §5.4; **M++** is treated in §5.5. For now the salient point is that the various naming schemes return CAs of different bit size (128 bits for **M** and 256 bits for **GM** and **M++**), and hence have differing collision probabilities and preimage resistance.

## 5.1 Birthday Paradox

Question: Ignoring leap years, how many people need to be present in a room for there to be ≈50% chance that someone in the room shares (i.e. collides with) your birthday? Answer: 253 (as shown in (5.1) below)[1].

$$1 - \left(\frac{364}{365}\right)^{253} \approx 0.50 \qquad (5.1)$$

However, if you change the question to include one key difference, the answer changes dramatically.

The modified question is: how many people need to be present in a room for *any two* people to have ≈50% chance of sharing the same birthday? Answer: 23. This number is surprisingly low, which leads to this phenomenon being given appellations such as "birthday surprise" or "birthday paradox." The surprisingly small result for the second question becomes clear when you consider that with 23 people in the room there are 253 *pairs* of people in the room; i.e., $23 \cdot 22 / 2 = 253$. Of course, for this result to hold the people in the room must have an even (random) distribution of birthdays; clearly this result would not stand if the room was full of twins.

We now add some rigor and generalization to the birthday paradox.

**Theorem 3:** Let $P_{collision}(N, q)$ denote the probability to have at least one collision when we throw $q > 1$ balls at random into $N > q$ buckets, giving us equation (5.2) below.

$$P_{collision}(N, q) \leq 0.5 \frac{q(q-1)}{N} \qquad (5.2)$$

In this paper we only present an upper bound to $P_{collision}(N, q)$ since that is relevant for our collision probability calculations. For a more exhaustive proof, including a lower bound, readers are referred to [26].

**Proof:** Let $C_i$ be the event that the $i$-th ball is thrown in a non-empty bucket, i.e. it collides with one of the previous balls. Then $P[C_i]$ is at most $(i-1)/N$ since when the $i$-th ball is thrown, there are at most $i-1$ different occupied buckets and the $i$-th ball is equally likely to land in any of them. Now we have,

---

[1] This maps to a preimage or second-preimage hash collision.

$$P_{collision}(N,q) = P[C_1 \vee C_2 \vee ... \vee C_q]$$

$$\leq P[C_1] + P[C_2] + ... + P[C_q]$$

$$\leq \frac{0}{N} + \frac{1}{N} + \frac{2}{N} + ... + \frac{q-1}{N} \quad (5.3)$$

$$\leq \frac{q(q-1)}{2N}.$$

From equation (5.3) we see the quadratic relation between $q$ and $P_{collision}(N,q)$. This is the core of the birthday paradox. You only need to throw about $\sqrt{N}$ balls into $N$ buckets to achieve a reasonable probability of ending up with two (or more) balls in the same bucket.

## 5.2 Centera with the M naming scheme

If a large number of messages are hashed using a given hash function, collisions will occur "by chance" after a very large number of messages have been hashed. Extensive statistical tests have been performed in the past on MD5 (for example in the RIPE project [19]), and no deviations from random behavior have been detected. This implies that modeling MD5 in this respect as a random function will result in a very good approximation of the collision probability.

MD5 hashes data into 128-bit strings. Relating to the proof in (5.3) we can state that $N = 2^{128}$. Therefore, once a threshold of $q = 2^{64}$ files is reached, collisions will be found very quickly. More precisely, if $q = \alpha \cdot 2^{64}$ files are hashed, then we can derive from formula (5.2) that the probability that one or more collisions occur is upper bounded by (5.4).

$$\frac{\alpha \cdot 2^{64} \cdot (\alpha \cdot 2^{64} - 1)}{2 \cdot 2^{128}} \approx \frac{\alpha^2}{2} \quad (5.4)$$

Note that Centera typically operates with very small values of $\alpha$ (e.g., .000000001) since the total file count in the system is substantially less than $2^{64}$ files (18 billion billion files).

Table 1 indicates the probability of a CA collision when using the **M** naming scheme for various fill rates.

| Number of objects | Capacity with avg file size 10 bytes | Capacity with avg file size 1K | M Collision Probability |
|---|---|---|---|
| 1.E+06 | 10 Megabyte | 1 Gigabyte | 1.E-27 |
| 1.E+07 | 100 Megabyte | 10 Gigabyte | 1.E-25 |
| 1.E+08 | 1 Gigabyte | 100 Gigabyte | 1.E-23 |
| 1.E+09 | 10 Gigabyte | 1 Terabyte | 1.E-21 |
| 1.E+10 | 100 Gigabyte | 10 Terabyte | 1.E-19 |
| 1.E+11 | 1 Terabyte | 100 Terabyte | 1.E-17 |
| 1.E+12 | 10 Terabyte | 1 Petabyte | 1.E-15 |
| 1.E+13 | 100 Terabyte | 10 Petabyte | 1.E-13 |
| 1.E+14 | 1 Petabyte | 100 Petabyte | 1.E-11 |
| 1.E+15 | 10 Petabyte | 1 Exabyte | 1.E-09 |

**Table 1. Collision Probability using the M naming scheme**

Such numbers are sufficiently large that it's difficult to get a sense of magnitude. Using time as the dimension, assume a hypothetical CAS application capable of storing 10,000 files per second to a Centera cluster. Using an average file size of 1K, writing $10^{15}$ such files would take up an exabyte of protected storage. However, it would take several millennia to fill up the cluster at this rate. As a point of comparison, after 1,000 years of continuous operation the probability of a CA collision using the M naming scheme is less than 1 in 1,000,000,000.

In the case of a malicious user seeking to forge two arbitrary files which return an identical **M** content address (this is referred to as "collision resistance"), then the user must attempt to apply the work described in [30][31] to Centera.

In the case of a malicious user seeking to forge a file X′ which gets an identical **M** content address as a given file X (this is referred to as "preimage resistance"), one can attempt to apply the ideas described in [11]. Knowing that the maximal file size in Centera is 100Mb, we have less than $2^{21}$ 512-bit blocks in one file. We can conclude that on average about $2^{128-21+1} = 2^{108}$ MD5 evaluations are required to forge a preimage of a given file X. Although this effort can happen offline, the number is sufficiently large to make such attacks impractical.

## 5.3 Random Generators

In order to move towards the stronger GM naming scheme, we need the following definitions from the HAC [2].

Definition 6: A **pseudorandom bit generator** (PRBG) is a deterministic algorithm which, given a truly random binary sequence of length $k$, outputs a binary sequence of length $l \geq k$ which "appears" to be random. The input to the PRBG is called the *seed*, while the output of the PRBG is called a *pseudorandom bit sequence*.

Definition 7: A pseudorandom bit generator is said to pass the **next-bit test** if there is no polynomial-time algorithm which, on input of the first $l$ bits of an output sequence $s$, can predict the $(l+1)^{st}$ bit of $s$ with probability significantly greater than ½.

Definition 8: A PRBG that passes the next-bit test is called a cryptographically secure pseudorandom bit generator (CSPRBG).

## 5.4 Centera with the GM naming scheme

Although the probabilities of having collisions with the M naming scheme are minimal, a second naming scheme, called GM, was introduced which includes additional random bits in the resultant CA. This additional randomness has the effect of further reducing the probability of collisions.

The Centera GM CA contains the following components:

- **M:** a 128 bit MD5 hash based on the file content;
- **G:** a 70 bit random string generated by a cryptographically secure pseudorandom bit generator;
- **T:** a 35 bit timestamp, with second resolution (actually, one unit of **T** is 1024 milliseconds);
- **C:** a 10 bit counter; and
- **H:** 13 bits used as "header" codes.

### 5.4.1 Timestamp and counter

The timestamp **T** significantly reduces the probability of collisions by effectively partitioning the entire Centera system into sets of files that have been written within about one second of each other. Incoming files are time stamped by the access node through which the file enters the cluster the first time. The timestamp **T** overflows after a little more than 1,000 years.

Suppose that the peak write speed to a Centera is $S$ files per second per access node. Given $A$ access nodes in a Centera, we conclude that the upper bound for the number of files having identical timestamp **T** is equal to $A \cdot S$.

Every access node in Centera maintains a 10-bit counter **C** which is increased for every new file stored on the system. The value of this counter **C** is randomized across the access nodes and repeats after each 1,024 files stored through the access node. An upper bound for the number of files having both identical timestamp **T** and identical counter **C** is (5.5).

$$A + \left\lfloor \frac{A \cdot S}{2^{10}} \right\rfloor \approx A + \frac{A \cdot S}{2^{10}} \qquad (5.5)$$

This formula holds for $S$ larger than $A$, which is always the case. For very large values of $S$, the addition of $A$ can be omitted as having minor impact on the result. Note that the $A$ term is due to a quantization effect because it is not possible to split documents into pieces for access node processing (e.g., when 1025 documents are assigned C values from the 1024 possibilities, at least one value will be assigned to at least 2 documents).

### 5.4.2 Applying the Birthday Paradox

In order for a collision to occur using the GM naming scheme, all components **M.G.T.C.H** of the blob address must be identical for two distinct files.

In §5.4.1 we showed that the impact of the timestamp **T** and counter **C** effectively limits the size of the set in which we can have collisions, to approximately $A \cdot S / 2^{10}$ files (assuming that $S$ is much larger than $A$).

The component **H** is non-random, hence it does not help to reduce the probability of collisions.

The components **M** and **G** are fully random and add up to 128+70=198 bits (see §5.2 for the randomness of **M** and §5.4 for the randomness of **G**).

Applying the birthday paradox to a set of $A \cdot S / 2^{10}$ balls thrown into $2^{198}$ buckets, we end up with the probability of a collision with the GM naming scheme indicated in (5.6).

$$\frac{(A \cdot S)^2}{2^{20} \cdot 2 \cdot 2^{198}} = \frac{(A \cdot S)^2}{2^{219}} \qquad (5.6)$$

The number of possible combinations (**T**, **C**) over a large time interval is equal to the number of milliseconds in that time interval.

If we run a Centera system for a period of $Z$ milliseconds, then for large values of $Z$, the probability of experiencing a collision during these $Z$ milliseconds is indicated in (5.7).

$$\left( \frac{(A \cdot S)^2}{2^{219}} \right) Z \qquad (5.7)$$

### 5.4.3 Collision Probability with GM

In a typical Centera setting, an upper bound for $A$ is 100 different access nodes, and an upper bound for $S$ is 10,000 small files per second per access node. Running this experiment for 1,000 years, we get $Z = 3.15 \cdot 10^{13}$ milliseconds. Feeding this into equation (5.7) we end up with the probability described in the equation (5.8) below.

$$\left( \frac{(100 \times 10,000)^2}{2^{219}} \right) Z \approx 4 \cdot 10^{-41} \qquad (5.8)$$

In short, the probability of a collision in the GM naming scheme is $\sim 4 \cdot 10^{-41}$ assuming an ingestion rate of 1,000,000 files/second over a period of 1,000 years. Note that this is an extraordinarily high ingestion rate over a long period of time. In the practical usage of Centera, the collision probability will be several orders of magnitude smaller.

To help put this into perspective, the probability of a CA collision is less likely than a bit error on disk (the non-recoverable Bit Error Rate (BER) of a typical ATA disk is ~1 for every $10^{15}$ bits read). As mentioned in §1, in addition to using the CA as an object handle, Centera uses the CA as an MDC. Therefore, in the event of a bit error on disk, Centera will detect this error via a background task that periodically checks the validity of a file image on disk against its CA. In the event a manipulation is detected, Centera will automatically replace the bad file with a good copy, thus insuring data integrity over the life of the archive.

In the case where a malicious user wishes to forge two arbitrary files which result in an identical **GM** content address (CA), the user needs to mount an online attack against a Centera system. One strategy would be to start from a known **M** collision (as described in [30]) and attempt to write the colliding file to all access nodes simultaneously within the same millisecond. In this way the attacker would seek to get some control over the M.T.C.H components of the **GM** content address (cf. §5.4). Within that millisecond, the user would attempt to write to *A* access nodes in parallel and hope for a collision in the 70-bit G component. Since *A* is typically very small (less than 100), coupled with the time it would take to perform the writes, it would take millions of years before such an attack is successful.

For the same reasons, forging a second preimage in an attempt to overwrite an existing **GM** file, is infeasible.

## 5.5 Centera with the M++ naming scheme

As discussed in §5.4, the **GM** naming scheme as implemented does not allow for single-instance store[2]. For those applications that would like to retain this ability, but use a stronger naming scheme than **M**, a third naming scheme was created. This 256-bit naming scheme, called **M++**, is constructed by the concatenation of 128 bits of MD5, 8 bits of formatting and 120 bits of SHA-256. The calculation of object names in this naming scheme requires more CPU cycles due to the introduction of the extra hash function evaluation.

Joux demonstrates that, contrary to Fact 9-27 in [2], the concatenation (or "cascading") of two iterated hash functions is only as strong as the strongest of the two hash functions [1]. The following results apply to the **M++** naming scheme.

---

[2] More precisely, GM – as presently implemented – does not allow identical objects to have identical names. The *function*, single-instance store, is anticipated to be activated for the GM naming scheme in a later release of CentraStar.

When writing **M++** objects to a Centera, collisions become likely after $2^{124}$ objects have been stored on the system (cf. the **M** naming scheme where this number is $2^{64}$). This can be validated by looking at formula (5.2) and substituting *q* with $2^{124}$ and *N* with $2^{120+128} = 2^{248}$ to reach a collision probability of ≈50%.

After 1,000 years of continuous operation, a hypothetical CAS application capable of storing 10,000 files per second on a Centera system would have stored $3.15 \cdot 10^{14}$ files. This results in a collision probability of $10^{-46}$ for the **M++** naming scheme (cf. $10^{-9}$ for the **M** naming scheme). This can be validated by looking at formula (5.2) and substituting *q* with $3.15 \cdot 10^{14}$ and *N* with $2^{120+128} = 2^{248}$.

A malicious user wishing to induce a collision for **M++** will need to calculate ~$2^{67}$ SHA-256 hash function evaluations. Potential future optimizations of the MD5 attacks described in [30] have no impact on this result. This result is achieved as follows. One can try to create a $2^{60}$-fold multicollision for MD5 using the techniques of Wang et al. [30] combined with the ideas of Joux [1]. This requires about 60 times more work than creating a single collision. The length of the resulting file will be at least 120 blocks of MD5, i.e. 7,680 bytes. Next, one searches within this space of $2^{60}$ files for a collision for the leftmost 120 bits of SHA-256. One expects to find such a collision by the birthday paradox. The effort required is $120 \cdot 2^{60} \approx 2^{67}$ evaluations of SHA-256.

In case a malicious user wants to forge a second preimage in an attempt to overwrite an existing **M++** file, we can use the ideas outlined in [11] to show that an effort of about $2^{119}$ SHA-256 evaluations is required.

## 5.6 Summary of naming schemes

The introduction of the **GM** and **M++** naming scheme illustrates the flexibility of the Centera architecture, specifically designed to allow for seamlessly switching amongst naming schemes. Cryptography will continue to evolve, and due to the high importance in the commercial and military world, hash functions will be under continuous scrutiny. Therefore, it is important that Centera is able to adapt its content addressing schemes to reflect the state of the art in cryptology.

Table 2 summarizes the strength of the various naming schemes.

The second column estimates the number of files that need to be written to a Centera system before collisions become likely.

The third column lists the effort required for an attacker to forge a collision, i.e. to forge two files X and X′, which result in the same content address on a Centera. Note that this is not the same as starting with a file X, and creating a second file X′ which has the same content address.

The fourth column lists the effort required for an attacker to forge a second preimage. For this the attacker starts with a file X and its content address, and the intent is to forge a second file X′ which gets the same content address on a Centera.

Note that for the **M** and **M++** naming schemes, an attacker could attempt to create a collision or a second preimage offline. However, for the **GM** naming scheme, all attacks need to happen online since the attacker has no control over the 70 bits random **G** part of the content address. This effectively makes attacks against **GM** infeasible, since it would take millions of years to execute these attacks.

|        | Number of files beyond which collisions become likely | Work required to forge file collision | Work required to forge $2^{nd}$ preimage of a given file X |
|--------|---|---|---|
| **M**  | $2^{64}$ files stored | $O(1)^3$ | $O(2^{108})$ |
| **GM** | Not possible | Not possible | Not possible |
| **M++**| $2^{124}$ files stored | $O(2^{67})$ | $2^{119}$ |

**Table 2. Summary of naming schemes**

## 6 Conclusion

In this paper we have described basic principles of operation for iterative hash functions in general and MD5 in particular. We then described the naming schemes used within Centera and mapped the properties described earlier to the Centera Content Address (CA). We demonstrated the likelihood of a CA collision in a Centera cluster. Finally, we concluded that hash functions will continue to evolve, and that the Centera system is architected to quickly adjust its naming scheme strategy to the latest insight in the relative strengths of cryptographic hash algorithms.

## 7 Acknowledgements

The authors would like to acknowledge Prof. Bart Preneel for his extensive help in writing this paper, and David Black for his help in deriving the proper function for computing collision probabilities when using the GM naming scheme.

We would also like to thank reviewers of early drafts of this paper for their time and useful comments. Specifically: Dirk Nerinckx, Jiri Schindler, Geert Denys, and Kristof Van Belleghem.

---

[3] For simplicity assume MD5 collisions are available from [30].